\documentclass [11 pt] {article}

\usepackage{amsmath}
\usepackage{amssymb}
\usepackage{amsthm}
\usepackage{epsfig}

\theoremstyle{plain}

\begin {document}

\title{On the time reversal invariance of classical electromagnetic theory\\
\vspace{-.5em}
{\normalsize [Version 3.3:   October 22, 2003]}}

\author{David B. Malament\thanks {\emph{E-mail address}:  dmalamen@uci.edu.} \\ 
{\small \emph{Department of Logic and Philosophy of Science,  
\vspace{-.5em}
University of California,}}  \\ 
{\small \emph{Irvine, CA  92697-5100, USA}}\\
{}\\
{\normalsize This article is dedicated to the memory of Rob Clifton.}}
\date{}
\maketitle
\vspace{-2em}
\begin{abstract}
\noindent David Albert claims that classical electromagnetic theory is not time reversal invariant. He acknowledges that all physics books say that it is, but claims they are ``simply wrong"  because they rely on an incorrect account of how the time reversal operator acts on magnetic fields.  On that account, electric fields are left intact by the operator,  but magnetic fields are inverted. Albert sees no reason for the asymmetric treatment, and  insists that \emph{neither} field should be inverted. I argue, to the contrary, that the inversion of magnetic fields makes good sense and is, in fact, forced by elementary geometric considerations.  I also suggest a way of thinking about the time reversal invariance of classical electromagnetic theory -- one that makes use of the invariant four-dimensional formulation of the theory -- that makes no reference to magnetic fields at all. It is  my hope that it will be of interest in its own  right, Albert aside.  It has the advantage that it allows for arbitrary curvature in the background spacetime structure, and is therefore suitable for the framework of general relativity. The only   assumption one needs is temporal orientability. 
\vspace{1em}

\noindent \emph{Keywords:}  time reversal invariance;  electromagnetic theory; relativity theory.
\end{abstract}

\section{Introduction}

In the first chapter of \emph{Time and Chance}, David Albert (2000) argues that  classical electromagnetic theory (in contrast, for example, to Newtonian mechanics)  is not time reversal invariant. He acknowledges that all physics books say that it is, but claims they are ``simply wrong"  because they rely on an incorrect account of how the time reversal operator, properly understood, acts on magnetic fields.  Once that account is corrected, he believes, it is perfectly obvious  that the theory is not time reversal invariant. No deep mathematics or physics is called for, only a clear understanding of the nature of the time reversal operation. 

Received opinion, no doubt, is often wrong.  But I don't believe it is here.   Physics  books tell us that the time reversal operation leaves the electric field \textbf{E}  intact, but inverts the magnetic field \textbf{B}.  Albert sees no reason for the asymmetric treatment, and  insists that \emph{neither} field should be inverted. He  even suggests (p. 18) that the inversion of \textbf{B} is nothing but an \emph{ad hoc} maneuver to save the time reversal invariance of classical electromagnetic theory. 
I'll argue, to the contrary (section 6), that the inversion of \textbf{B} makes good sense and is, in fact, forced by elementary geometric considerations.   The argument -- really just a version of one that can be found in any book on the subject -- traces the asymmetric treatment of \textbf{E} and \textbf{B} to the fact  that the latter, unlike the former, is not a vector field in the usual sense. (In traditional language,  \textbf{E} is a ``polar" vector field, while \textbf{B}  is an ``axial" vector field.) 

Before giving this response to Albert's claims, I'll make a somewhat different point. It seems to me that the inversion of magnetic fields by the time reversal operator is really something of a distraction.  One can formulate and argue for the claim that classical electromagnetic theory is time reversal invariant without making reference to magnetic fields at all.  I'll do so in sections 3 and 4, using the invariant four-dimensional formulation of the theory. It is my hope that the proposed way of thinking about time reversal invariance will be of interest in its own right, Albert aside. 

The key idea is this. The tensor fields  $(F_{ab}, J^a)$ that represent the electromagnetic field and its charge-current source field are only determined relative to a choice of temporal orientation. I'll construe time reversal as an operation taking pairs $(F_{ab}, J^a)$ as determined relative to one orientation to pairs $({}^T\hspace{-.2em}F_{ab}, {}^T\hspace{-.2em}J^a)$ as determined relative to the other.  This approach has the advantage that it allows for arbitrary curvature in the background spacetime structure, and is therefore suitable for the framework of general relativity.  The only assumption one needs is temporal  orientability. (In contrast, the standard approach presupposes that the background spacetime structure exhibits special time reflection symmetries.\footnote{The standard approach leaves the background temporal orientation fixed, but inverts dynamical histories under the action of the symmetries.})  At the same time, it is fully equivalent to the standard approach when the symmetries \emph{are} present, as in Minkowski  spacetime.\footnote{My  discussion of Albert's views is closely related to those of John Earman (2002) and Frank Arntzenius (2003/4?). They too dispute his claims about the (non) temporal invariance of classical electromagnetic theory,  but offer somewhat different arguments in response.}

\section{Albert's Argument}

I'll start by presenting the standard account of time reversal invariance and then reconstructing Albert's argument. 

The standard account goes something like this.\footnote{I will, for the moment, take for granted that we have a well defined notion of space at a given time and ignore complications involving relativity theory.}  A physical theory is said to be  \emph{time reversal invariant} if, for any sequence of instantaneous states  $S_I, ...,  S_F$ allowed by the theory,  the time reversed sequence $ {}^R(S_F), ...,  {}^R(S_I)$ is allowed as well. (Here $R$ is the \emph{time reversal operator}, and the temporal order of states is understood to run from left to right.) If a time coordinate $t$ is given, we can formulate the defining condition, somewhat more precisely, this way:   for any history $t \mapsto S(t)$  allowed by the theory, the time reversed history  
$t \mapsto ({}^T\hspace{-.2em}S)(t)$ is allowed as well, where $({}^T\hspace{-.2em}S)(t) = {}^R(S(-t))$.\footnote{The notation may be confusing here. I am taking $R$ to be an operator acting on individual instantaneous \emph{states}, and taking $T$ to be an operator acting on \emph{histories} that is induced by $R$. The ``time reversed history" $t \mapsto ({}^T\hspace{-.2em}S)(t)$  runs the states of  $t \mapsto S(t)$ in reverse temporal order \emph{and} applies $R$ to each one.}  For these characterizations to make full sense in any particular case, of course, we have to know what count as ``instantaneous states",  and how the time reversal operator  $R$  acts on them.

Consider, for example,  the case of a point particle in Newtonian mechanics.  Here (on the standard account) the instantaneous states are pairs  $(\textbf{x}, \textbf{v})$, where $\textbf{x}$ is the particle's position and $\textbf{v}$ its velocity; and the time reversal operator $R$ takes the state $(\textbf{x}, \,  \textbf{v})$ to the state $(\textbf{x}, \,  -\textbf{v})$. It follows that  the  induced operator $T$ takes the history 
$t \mapsto S(t) =(\textbf{x}(t),\, \textbf{v}(t))$ to the time reversed history
\[
t  \mapsto ({}^T\hspace{-.2em}S)(t) = {}^R(\textbf{x}(-t),\, \textbf{v}(-t)) = (\textbf{x}(-t),\, -\textbf{v}(-t)).
\]
The latter reverses the  order in which the particle occupies particular positions, and inverts its velocity at every one.  The latter inversion (turning $\textbf{v}$ to $-\textbf{v}$) makes sense. If we watch a movie of a particle moving from left to right, and then run the movie  backwards, we see it moving from right to left.  Since the velocity of a particle is the time derivative of its position, $R$ \emph{must} invert $\textbf{v}$.\footnote{Since the position of the particle in the time reversed trajectory is  $({}^T\textbf{x})(t) = \textbf{x}(-t)$ at  time t, its velocity at t is
\[
({}^T\textbf{v})(t) = \frac{d}{dt} \ ({}^T\textbf{x})(t) = \frac{d}{dt} \ \textbf{x}(-t) = -\textbf{v}(-t).
\]
}

Next consider (the standard account of) classical electromagnetic theory. Here the instantaneous states are quadruples $(\textbf{E}, \, \textbf{B}, \rho, \, \textbf{j})$, where $\textbf{E}$ is the electric field, $\textbf{B}$ is the magnetic field, $\rho$ is the charge density field, and $\textbf{j}$ is the current density field. The latter two characterize the instantaneous state of the charge distribution that serves as a source for $\textbf{E}$ and $\textbf{B}$. 
The time reversal operator $R$ (at least according to the standard account) has the following action on these objects:
\begin{equation} \label{RonEand B}
(\textbf{E}, \ \textbf{B}, \ \rho, \ \textbf{j})  \hspace{2em}   \stackrel{R}{\longmapsto}
\hspace{2em}  (\textbf{E}, \  -\textbf{B},  \  \rho, \  -\textbf{j}).
\end{equation}
Hence the induced operator $T$ takes the composite history 
\begin{equation} \label{originalhistory}
t \mapsto (\textbf{E}(t,\textbf{x}), \ \textbf{B}(t, \textbf{x}), \  \rho(t, \textbf{x}),  \ \textbf{j}(t, \textbf{x}))
\end{equation}
to the time reversed history
\begin{equation} \label{timereversedhistory}
t \mapsto \left( ({}^T\textbf{E})(t,\textbf{x}), \ ({}^T\textbf{B})(t, \textbf{x}), \  ({}^T\rho)(t, \textbf{x}),  \ ({}^T\textbf{j})(t, \textbf{x}) \right),
\end{equation}
where
\vspace{-1em}
%\begin{center}
%\emph{Standard time reversal transformations}
%\end{center}
%\vspace{-1.5em}
\begin{eqnarray}
({}^T\textbf{E})(t, \textbf{x}) &=& \textbf{E}(-t, \textbf{x}),     \label{STRT1}  \\ 
({}^T\textbf{B})(t, \textbf{x}) &=&  -\textbf{B}(-t, \textbf{x}),    \label{STRT2}  \\
({}^T\rho)(t, \textbf{x})  &=& \rho(-t, \textbf{x}),   \label{STRT3}  \\ 
({}^T\textbf{j})(t, \textbf{x}) &=& -\textbf{j}(-t, \textbf{x}).   \label{STRT4}   
\end{eqnarray}
These are the \emph{standard time reversal transformations}. (Here I make explicit the dependence of the fields on position; $\textbf{E}(t,\textbf{x})$, for example,  is the value of the electric field at time t and position $\textbf{x}$.)   Just why (or whether) $R$ should have the action indicated in (\ref{RonEand B}) -- in particular, why it should invert the magnetic field $\textbf{B}$ but leave intact the electric field $\textbf{E}$ -- is precisely the issue I will be considering. (Albert  claims  it should  \emph{not} invert $\textbf{B}$.)  But if $R$  \emph{does} act as in (\ref{RonEand B}), then it follows immediately that the original history  (\ref{originalhistory}) satisfies Maxwell's equations iff the time reversed history  (\ref{timereversedhistory}) does so, i.e., it follows that classical electromagnetic theory (or, at least, that fragment of the theory embodied in these equations) is time reversal invariant. This can be seen by inspection of Maxwell's equations:
%\vspace{-.5em}
%\begin{center}
%\emph{Maxwell's equations}
%\end{center}
%\vspace{-1.em}
\begin{eqnarray}
\nabla   \cdot \textbf{E}  & = &  \rho,    \label{1st.3dim.Maxwell}  \\
\nabla   \times \textbf{E}  & = &  - \frac{ \partial \textbf{B}}{ \partial t},   \label{2nd.3dim.Maxwell} \\
\nabla   \cdot \textbf{B}  & = &  0       \label{3rd.3dim.Maxwell},  \\
\nabla  \times \textbf{B}  & = &  \frac{ \partial \textbf{E}}{ \partial t}  + \textbf{j}.      \label{4th.3dim.Maxwell}
\end{eqnarray}

\vspace{1em}

David Albert's objection to the standard account of time reversal invariance begins with a disagreement over how to characterize instantaneous states. Consider again the example of a particle in Newtonian mechanics. It is one of Albert's fundamental claims that though the particle's \emph{position} at a given moment (time $t_0$) deserves to be considered a component of its instantaneous state, its \emph{velocity} at that moment  does  not. He reaches this  conclusion by first insisting that (on any proper account)  (i) instantaneous state descriptions  at different times  should be ``logically, conceptually, and metaphysically independent" of each other (p. 10), and then pointing out that (ii)  this requirement fails if we characterize instantaneous particles states in terms of positions \emph{and} velocities. It fails because specification of the particle's position at all times before and after $t_0$ \emph{determines} its velocity at $t_0$. As he  puts it: 

\begin{quote}
\small{
What typically gets referred to in the \emph{physical literature} as an ``instantaneous state" of a Newtonian-mechanical universe, of course, is a specification of the positions \emph{and the velocities} of all the particles in the world at the time in question. But the trouble with that is just that specifications of the positions \emph{and the velocities} of all the particles in the world at one time are \emph{not} conceptually \emph{independent} of specifications of the positions and velocities of all the particles in the world at all \emph{other} times. The trouble (to put it slightly differently) is that a specification of the positions and velocities of all the particles in the world at some particular instant is \emph{not} a specification of the physical situation of the world at that instant \emph{alone}; it is \emph{not} a specification of the physical situation of the world at that instant \emph{as opposed to all others}, at all! \  (Albert, 2000, pp. 10-11,   italics in original)}
\end{quote}

I am not convinced by Albert's criticism of the standard account of instantaneous states. Responses by Earman (2002), Floyd (2003), and Smith (2003) seem to me exactly right.\footnote{The papers by Floyd and Smith are directed primarily against Arntzenius (2000).  But  most of what they have to say applies equally well  to Albert (2000).} But I  am not going to dwell on the criticism because it is largely irrelevant to my concerns. Albert believes that the things called ``instantaneous states"  in  physics books should properly be called  ``dynamical  conditions".  Fine. Suppose we adopt that terminology. Then the standard characterization of time reversal invariance given above comes out as follows:  a physical theory is time reversal invariant if, for any sequence of dynamical conditions  $S_I, ...,  S_F$ allowed by the theory,  the time reversed sequence $ {}^R(S_F), ...,  {}^R(S_I)$ is allowed as well.  This is not Albert's preferred characterization\footnote{He takes a physical theory to be ``time reversal invariant" if, for any sequence of  ``Albert instantaneous states" $S_I, ...,  S_F$ allowed by the theory,  the sequence $ S_F, ...,  S_I$ is allowed as well.  (Here ``Albert instantaneous states" are ones that satisfy his strictures. In the case of Newtonian mechanics, they include particle positions, but not velocities.)}, but, as he acknowledges (p. 19), it is equivalent if one adopts what he takes to be the proper account of the action of the time reversal operator $R$. So I am going to put other issues (such as the nature of ``instantaneous states") aside, and focus my attention specifically on his non-standard account of that operator. That seems to me the heart of the matter.

%and construe his dissent from the standard account of time reversal invariance as one specifically concerning the action of that operator 

%and focus my attention specifically on his non-standard account of that operator.   

%inclined to put other issues (such as the nature of ``instantaneous states") aside and construe his dissent from the standard account of time reversal invariance as one specifically concerning the action of that operator.  

% it seems to me that \emph{the heart of the matter is not what terminology one employs, but rather how one understands the action of that operator} (and those are the terms in which  I will reconstruct his position).

%Albert does not like this characterization, but he acknowledges (p. 19)  that it is equivalent to his preferred alternative \emph{if} one adopts what he takes to be the proper account of the action of the time reversal operator R.  
%So it seems to me that \emph{the heart of the matter is not what terminology one employs, but rather how one understands the action of that operator} (and those are the terms in which  I will reconstruct his position).

Albert agrees that if  we have two vector quantities in a basic dynamical condition, and the second is the time derivative of the first, then if $R$ keeps the first intact,  it must reverse the second. (The argument is essentially as in footnote 5.)   So he agrees that in the case of Newtonian particle mechanics, $R$ takes the dynamical  condition $(\textbf{x}, \textbf{v})$ to $(\textbf{x}, -\textbf{v})$. So he agrees that the theory is time reversal invariant (despite his disagreement as to whether a particle's velocity counts as part of its ``instantaneous state"). \emph{Where he takes his stand is with the principle that $R$ should not reverse a vector quantity in a basic dynamical condition \emph{unless} it arises as the time derivative of some other quantity that $R$ leaves intact.} That is why he does not accept (\ref{RonEand B}) as a characterization of $R$'s action in classical electromagnetic theory. Specifically, he insists  that just as $R$ leaves $\textbf{E}$ intact, it should leave $\textbf{B}$ intact.

\begin{quote}
\begin{small}
Magnetic fields are \emph{not} the sorts of things that any proper time-reversal transformation can possibly turn around. Magnetic fields are not -- either logically or  conceptually-- the \emph{rates of change} of anything.  (Albert, 2000, p. 20, italics in original)
\end{small}
\end{quote}

\noindent He takes the proper action of $R$ to be, not (\ref{RonEand B}), but rather\footnote{Albert does not discuss sources. I simply take for granted that he would have $R$ reverse the current density $\textbf{j}$ since it is a velocity-like object. But nothing turns on this assumption. I could equally well restrict attention to Maxwell's equations in the special case where all sources vanish: $\rho = 0 = \textbf{j}$. In that case too, the equations (as a set) are invariant under the standard time reversal transformations, but not under Albert's proposed alternative transformations.}

\vspace{-1em}

\begin{equation} \label{alternateRonEand B}
(\textbf{E}, \ \textbf{B}, \ \rho, \ \textbf{j})  \hspace{2em}   \stackrel{R \ (Albert \  alternative)}{\longmapsto}
\hspace{2em}  (\textbf{E}, \  \textbf{B},  \  \rho, \  -\textbf{j}).  
\end{equation}

\noindent It follows that, for him,  the induced operator $T$ takes the composite history 
\begin{equation*} 
t \mapsto (\textbf{E}(t,\textbf{x}), \ \textbf{B}(t, \textbf{x}), \  \rho(t, \textbf{x}),  \ \textbf{j}(t, \textbf{x}))
\end{equation*}
to the time reversed history
\begin{equation*} 
t \mapsto \left( ({}^T\textbf{E})(t,\textbf{x}), \ ({}^T\textbf{B})(t, \textbf{x}), \  ({}^T\rho)(t, \textbf{x}),  \ ({}^T\textbf{j})(t, \textbf{x}) \right),
\end{equation*}
where, now, the terms on the right are given by the following table of \emph{Albert's proposed time reversal transformations}:
%\vspace{-1.em}
\begin{eqnarray}
({}^T\textbf{E})(t, \textbf{x}) &=& \textbf{E}(-t, \textbf{x}),    \label{ATRT1}   \\ 
({}^T\textbf{B})(t, \textbf{x}) &=&  \textbf{B}(-t, \textbf{x}),  \label{ATRT2} \\
({}^T\rho)(t, \textbf{x})  &=& \rho(-t, \textbf{x}),   \label{ATRT3}  \\ 
({}^T\textbf{j})(t, \textbf{x}) &=& -\textbf{j}(-t, \textbf{x}).  \label{ATRT4}   
\end{eqnarray}

\noindent The only difference between the standard and alternate transformations is in the treatment of $\textbf{B}$, i.e., in the difference between (\ref{STRT2}) and (\ref{ATRT2}). But everything turns on that difference. If Albert is correct that the  time reversal operator, properly understood, does not invert  $\textbf{B}$, then classical electromagnetic theory does not qualify as time reversal invariant.  (Two of Maxwell's equations are invariant under his alternate transformations. But the other  two -- the second and the fourth in our list -- are not.) 

\begin{figure} [h]
\begin{center}
\setlength{\unitlength}{1cm}
\begin{picture}(9.54,6.1)
  \put(0,0.0){\epsfig{figure=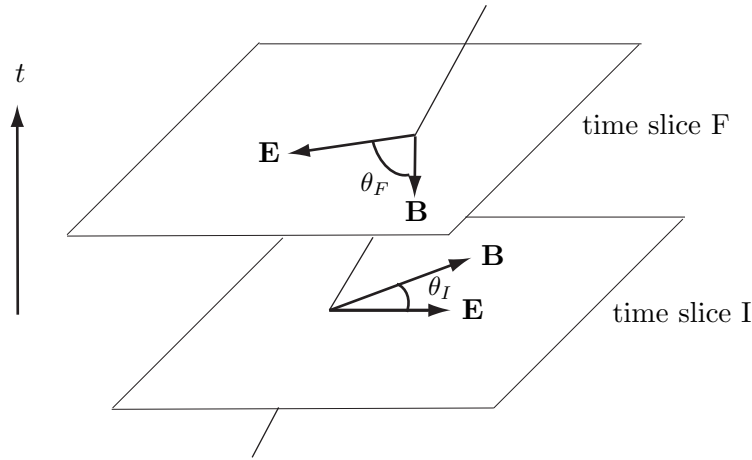}}
   \put(.06,5.0){\normalsize $t$}
   \put(8.,1.8){\normalsize time   slice I}
   \put(7.6,4.3){\normalsize  time   slice F}
   \put(6.0,1.84){\normalsize \textbf{E}}
    \put(6.255,2.57){\normalsize \textbf{B}}
   \put(3.29,3.93){\normalsize \textbf{E}}
    \put(5.24,3.15){\normalsize \textbf{B}}
   \put(5.55,2.165){\small $\theta_I$}
   \put(4.65,3.55){\small $\theta_F$}
 \end{picture} 
%\hspace{1em}
 \begin{minipage}[b]{14.0cm}

\caption{A partial history showing the \textbf{E} and \textbf{B} fields at two points on the worldline of a particle moving from left to right.} \label {E-Bhistory}
\end{minipage}
\end{center}
\end{figure}

My reconstruction of Albert's position to this point has been rather formal. So let me also try to capture what I take to be the intuitive picture behind it. I'll call it the ``$\textbf{E}$ and $\textbf{B}$  just lie there picture".  Consider a stacked family of simultaneity slices, with the bottom slice labeled `I' (initial)  and the top one labeled `F' (final). (See Fig. \ref{E-Bhistory}.) 
On each slice, two vectors fields,  $\textbf{E}$ and $\textbf{B}$ are given. Each can be pictured as an array of arrows, one at each point of the slice.  Consider a particle moving with uniform velocity ``from left to right". At each point where its worldline intersects a timeslice, the $\textbf{E}$ and $\textbf{B}$  vectors  have particular lengths and directions. Suppose that at  all these points the vectors are non-zero,  so that there is a well-defined angle between them. Let it be $\theta_I$  at the intersection with slice I, and $\theta_F$ at the intersection with slice F. 
 
\begin{figure}[h]
\begin{center}
\setlength{\unitlength}{1cm}
\begin{picture}(9.0,6.3)
  \put(0,0.0){\epsfig{figure=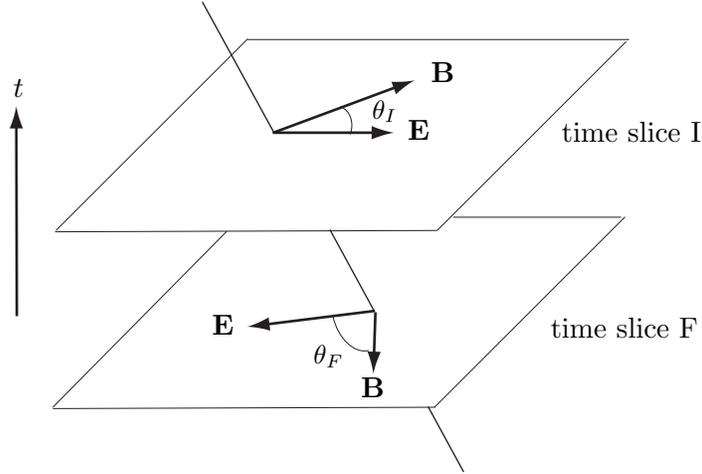}}
   \put(.06,5.0){\normalsize $t$}
  \put(7.2,1.79){\normalsize time   slice F}
   \put(7.35,4.4){\normalsize  time   slice I}
   \put(5.3,4.42){\normalsize \textbf{E}}
   \put(5.61,5.19){\normalsize \textbf{B}}
  \put(2.7,1.83){\normalsize \textbf{E}}
  \put(4.68,1.0){\normalsize \textbf{B}}
  \put(4.83,4.71){\small $\theta_I$}
  \put(4.05,1.45){\small $\theta_F$}
 \end{picture} 
%\hspace{1em}
 \begin{minipage}[b]{14.0cm}

\caption{The ``Albert time reversal" of the partial history displayed in Fig. \ref {E-Bhistory}. The order of the time slices I and F has been reversed, and the worldline is now that of a particle moving from right  to left.  But the configuration of  \textbf{E} and \textbf{B} fields within each slice has been left intact.} \label {E-BAlbertreversedhistory}
\end{minipage}
\end{center}
\end{figure}

Now think of each slice as a frame in a movie, and run the movie backwards. The result corresponds to an inverted stack of simultaneity slices, with slice F at the bottom and slice I at the the top.  (See Fig. \ref {E-BAlbertreversedhistory}.)  The order of the slices changes, and the particle is now seen to move ``from right to left".  But -- so the argument goes -- the configuration of  $\textbf{E}$ and $\textbf{B}$ vectors \emph{within} each slice does not change.  In particular, the $\textbf{E}$ and $\textbf{B}$ vectors at each point on the particle worldline are as before. So the angles between them are as before.   In the original movie, we see the angle $\theta_I$ at the beginning, and see $\theta_F$ at the end.  When the movie is run backwards, the order is reversed.  We first see $\theta_F$ and then later $\theta_I$.  (We do \emph{not} see $(\pi - \theta_F)$ followed by $(\pi - \theta_I)$, as we should if the $\textbf{B}$ vectors had been inverted in each frame before the movie was run backwards.)  

In this picture, the vector fields $\textbf{E}$ and $\textbf{B}$ in each slice ``just lie there" and are kept intact when the order of time slices is reversed.  I will eventually try to make the case that the picture is misleading. Roughly speaking, the idea will be that, because $\textbf{B}$ is an ``axial" vector field, it can no more just lie there than an angular velocity vector field can;  and the inversion of $\textbf{B}$ under time reversal is no more mysterious than  the change from clockwise rotation to counterclockwise rotation induced by that operation.  But first I will switch attention to the four-dimensional invariant formulation of classical electromagnetic theory, and give a direct argument for the time reversal invariance of the theory that makes no reference to magnetic fields at all. 

\section{The Invariant Formulation of Classical Electromagnetic Theory}

In this section, I review the invariant formulation of classical electromagnetic theory within the framework of relativity theory.\footnote{I will assume familiarity with the basic elements of relativistic spacetime geometry, and review just a few points of special importance. A good reference for all the material in this section and those that follow, including the abstract index notation that I  will be using, is Wald (1984).}

In what follows, let $(M, g_{ab})$ be a \emph{relativistic spacetime}, i.e.,  let $M$ be a smooth, connected, four-dimensional manifold, and let $g_{ab}$  be a smooth pseudo-Riemannian metric  on $M$ with signature $(1,3)$. (With this sign convention, a vector $\xi^a$ at a point of $M$ qualifies as \emph{timelike} if  $\xi^a \xi_a > 0$ and \emph{spacelike} if $\xi^a \xi_a  < 0$. It is \emph{null} if $\xi^a \xi_a = 0$.)  Furthermore,  let $\nabla_a$ be the unique (torsion free) derivative operator on $M$ compatible with $g_{ab}$  (i.e., $\nabla_a \, g_{bc}  =  0$). Finally, let $\tau^a$ be a continuous timelike vector field on $M$.\footnote{Thus I am assuming that $(M, g_{ab})$ is \emph{temporally orientable}, i.e., that there \emph{exists} such a field on $M$. (Note that timelike vectors are automatically non-vanishing since the zero vector at every point is null.)}

$\tau^a$ determines a ``temporal orientation" on $M$. I will say that a timelike vector $\xi^a$ at a point is \emph{future-directed}  relative to $\tau^a$ if  $\tau^a \xi_a  >  0$ and \emph{past-directed} relative to $\tau^a$ if  $\tau^a \xi_a  <  0$. (The first condition just captures the requirement that $\xi^a$ and $\tau^a$ be co-oriented at the point, i.e., that they belong to the same lobe of the null cone, and the second that they belong to opposite lobes.\footnote{Note that all  timelike vectors at all points qualify as \emph{either} future-directed or past-directed relative to $\tau^a$, since it is not possible that $\tau^a \xi_a = 0$. No two timelike vectors can be orthogonal.})  In what follows, we will need to keep track of the role played by this temporal orientation in our definitions and constructions.

In the invariant formulation of electromagnetic theory,  one works with a smooth, anti-symmetric tensor field  $F_{ab}$ on $M$ that represents the electromagnetic field itself, and a smooth vector field $J^a$ on $M$ that represents the charge-current field that serves as its source. Consider first $F_{ab}$.  We can think of it as coding, for every point in spacetime,  the electromagnetic force\footnote{The force experienced by a charged particle in the presence of an electromagnetic field can be decomposed into separate ``electric" and ``magnetic" components -- we will see how this works later -- but here I am considering only the net, undifferentiated force.} that would be experienced by a point test particle there, depending on its charge and instantaneous velocity.\footnote{A ``test particle" is one whose own contribution to the electromagnetic field is negligible and may, at least for the purposes at hand, be ignored.} But just how it does so is a bit delicate because of a certain ambiguity in the notion of ``instantaneous velocity".  One must distinguish between senses that do and do not depend on a choice of temporal orientation.

\begin{figure}
\begin{center}
\setlength{\unitlength}{1cm}
%\begin{picture}(12.6,6.0)
\begin{picture}(1.9,5.5) 
  \put(0.0,0.0){\epsfig{figure=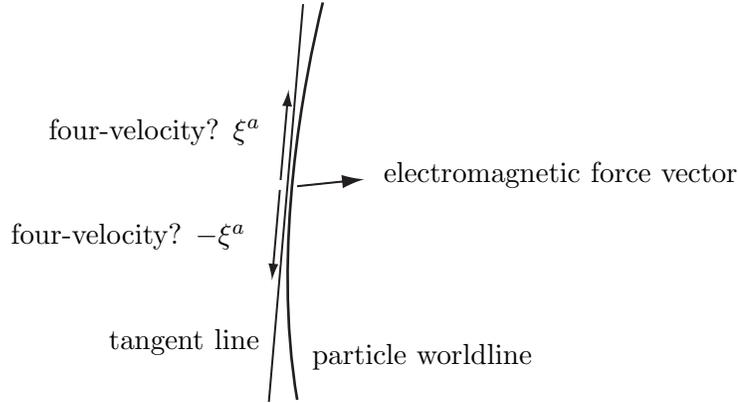}}
  \put(1.55,2.92){electromagnetic force vector}
 \put(.6,.5){particle worldline}
  \put(-2.9, 3.5){\normalsize four-velocity?  $\xi^a$}
  \put(-3.4, 2.1){\normalsize four-velocity?  $-\xi^a$}
   \put(-2.1, .7){\normalsize tangent line}
  % \put(.2, .8){\normalsize (or invariant ``instantaneous velocity" line)}
\end{picture} 
 \begin{minipage}[b]{14.0cm}

\caption{The ``instantaneous velocity" of a point particle at a point can be represented either as a tangent line, or as a four-velocity vector, there.  The latter depends on a choice of temporal orientation. The former does not.} \label {tangent.vectors}
\end{minipage}
\end{center}
\end{figure}

Consider a point particle with  non-zero mass.\footnote{Here and in what follows, ``mass" will be understood to mean rest mass.}  We can represent its worldline as the image of a smooth timelike curve  in $M$,  and take its tangent line at any point (in the tangent space of the point) to represent its instantaneous velocity there. (See Fig. \ref{tangent.vectors}.)   It is a basic fact of electromagnetic life that the net electromagnetic force experienced by the particle at a point depends only on its tangent line $L$ there and its charge q.
%\footnote{The fact in question has direct, experimental significance. We can build devices that measure the net electromagnetic force acting on a particle and display it as a vector displacement.  
%(We can, for example, rig the particle in a device with plastic (electrically neutral) springs and observe its displacement from some initial baseline point.) 
%If we make a movie of the measurement procedure, and run the movie  backwards,  then, when the critical moment arrives, we see the same vector displacement  that we saw the first time through.} 
%(Essentially the same story  serves to justify experimentally the claim that electric fields (as determined relative to frames of reference) are left intact under time reversal.  The net electromagnetic force vector experienced by a test particle carrying unit charge at a point just \emph{is} what we would otherwise describe as the electric field vector at the point  as determined relative to  a frame of reference in which the particle is instantaneously at rest .} 
So we have a force map of form  $<\hspace{-.4em}L, \, q\hspace{-.4em}> \hspace{.2em} \mapsto F(L, q)$  at every point.\footnote{Here I have temporarily dropped tensor indices. I hope this does not cause confusion. I could could equally well have written: $<\hspace{-.4em}L, \, q\hspace{-.4em}> \hspace{.2em} \mapsto F^a(L, q)$.} It is crucially important for our purposes that \emph{this structural characterization of the electromagnetic field makes no reference to the background temporal orientation}.

But when it comes to actually specifying the map, it is convenient to allow such reference. One would like to say, for example, that, for given $q$, the induced map  $L \mapsto F(L, q)$ is ``linear".  But it must be explained what that means,  since we are dealing here with a map from lines to vectors.  
The standard way is to redescribe the map as one from vectors to vectors using the background temporal orientation as an auxiliary structure. Rather than representing the particle's instantaneous velocity at a point as a tangent line, we represent it as a unit timelike vector there (co-aligned with the tangent line). But there are two from which to choose. (See Fig. \ref{tangent.vectors}.)  One is future-directed with respect to $\tau^a$; the other is past-directed.  It makes no difference which we choose, but, at least for the moment, let us systematically (i.e., at every point) work with the one that is future-directed.  
%This amounts to choosing the standard \emph{four-velocity} vector of the particle at a point (as determined relative to $\tau^a$) to represent its instantaneous velocity there.

So far, we have recast the force map at every point as one of form $<\hspace{-.2em}\xi^a,  q\hspace{-.2em}> \hspace{.2em} \mapsto F(\xi^a, q)$, where $\xi^a$ is a four-velocity vector, i.e., a future-directed, unit timelike vector, there.\footnote{Until further notice, let us agree that all references to temporal orientation, four-velocities, etc. should be understood to be determined relative to $\tau^a$.} Now we can avail ourselves of the mathematical formalism of linear algebra to recover the standard representation of the electromagnetic field.  It is another basic fact of electromagnetic life that the force map under consideration satisfies the following three simple conditions at every point. 
\begin{enumerate}
  \item [(i)]  For all $q$, the induced map  $\xi^a \mapsto F(\xi^a, q)$ is linear. 
  \item  [(ii)] For all four-velocity vectors $\xi^a$ and all $q$,  the force vector $F(\xi^a, q)$ is orthogonal to $\xi^a$. 
  \item  [(iii)] For all four-velocity vectors $\xi^a$, $F(\xi^a, q)$ is directly proportional to $q$ in the sense that   
  \[  
  F(\xi^a, q_1) = \frac{q_1}{q_2} \ F(\xi^a, q_2)
  \] 
 for all $q_1$ and all $q_2 \neq 0$.
\end{enumerate}
The three conditions (jointly) are equivalent to the requirement that there exist an anti-symmetric tensor field $F_{ab}$ on $M$ such that, at every point, 
\begin{equation}  \label{form of force}
F(\xi^a, q) \ = \ q \, F^{a}_ {\ b} \, \xi^b.\footnote{The argument proceeds in three steps. It follows from (i) that, for each $q$, there is a tensor field $\stackrel{q}{F}_{ab}$ on $M$ such that, at every point, $F(\xi^a, q) \ = \  \stackrel{q}{F}{}\hspace{-.1em}^{a}_ {\ b} \, \xi^b$ for all four-velocity vectors $\xi^a$. If we now take $F_{ab}$  to be $\stackrel{1}{F}_{ab}$, it  follows immediately from (iii) that
\[
  F(\xi^a, q) = \frac{q}{1} \ F(\xi^a, 1) = q  \stackrel{1}{F}{}\hspace{-.1em}^{a}_ {\ b} \, \xi^b =  q \, F^{a}_ {\ b} \, \xi^b.
\]
Now, finally, condition (ii) implies that, at all points, $ F_{ab} \, \xi^a  \xi^b =  (F^{a}_ {\ b} \, \xi^b) \, \xi_a = \textbf{0}$  for all four-velocity vectors $ \xi^a$.  Since the tangent space at a point is spanned by the set of four-velocity vectors there, it follows that $F_{ab} = -F_{ba}$ everywhere. (The converse holds too, for if $F_{ab}$ is an anti-symmetric tensor  field on $M$, and equation (\ref{form of force}) holds, then the force field $<\hspace{-.2em}\xi^a,  q\hspace{-.2em}> \hspace{.2em} \mapsto F(\xi^a, q)$ certainly satisfies conditions (i), (ii), and (iii) at every point.) }  
\end{equation}
In this way we are led to the field $F_{ab}$ with which we started.\footnote{Note that there can be only one tensor field $F_{ab}$ on $M$ satisfying (\ref{form of force}) at all points. For if two fields $\overset{1}{F}_{ab}$ and $\overset{2}{F}_{ab}$ code the same force values at all points, i.e., if  $(\overset{1}{F}\,^{a}_ {\ b} \,  - \overset{2}{F}\,^{a}_ {\ b}) \, \xi^b = 0$ for all four-velocity vectors $\xi^a $ at all points, then $\overset{1}{F}_{ab} =  \overset{2}{F}_{ab}$. This follows because the set of four-velocity vectors at any point of $M$ spans the tangent space there.} 

In our characterization of $F_{ab}$, we made essential reference to the background temporal orientation determined by $\tau^a$. At a crucial stage, we replaced tangent lines by unit tangent vectors, future-directed with respect to $\tau^a$. Schematically, we have:  
 
\[
\left.
\begin{array} {lc}
{}\hspace{4em}&\textrm{invariant map from}  \\
{}&\textrm{tangent lines to force vectors}   \\
{}& +  \\
{}&\textrm{temporal orientation}  
\end {array}
 \right\}  \hspace{1em} 
 \Longrightarrow \hspace{1em}    F_{ab} 
 \]   \vspace{0em}
 
\noindent But it should be stressed, once again, that the electromagnetic field itself can be fully characterized without reference to temporal orientation (as  a map from tangent lines to force vectors).

A similar story can be told about the second element in our initial pair $(F_{ab}, J^a)$, the vector field $J^a$ on $M$ that represents the charge-current source for the electromagnetic field. It is another basic fact of electromagnetic life that the (scalar) charge density observed at any point by an individual is determined solely by his  or her tangent line $L$ there. So we have a map of structure $L \mapsto J(L)$ at every point. If we systematically replace tangent lines $L$ with four-velocity vectors $\xi^a$, the map can be recast in the  form $\xi^a \mapsto J(\xi^a)$, i.e., as one from vectors to scalars (rather than from lines to scalars).  The map, it turns out, is linear. (This is the last basic fact of electromagnetic life I will need to cite for the characterization of $F_{ab}$ and $J^a$.) Linearity implies (and is fully equivalent to the requirement) that there is a vector field $J^a$ on $M$ such that, at every point,
\begin{equation}  \label{form of charge}
J(\xi^a) \ = \ J^{a} \, \xi_a
\end{equation}
for all four-velocity vectors $\xi^a$.\footnote{Clearly, there can be only one such field.  The argument is much the same as given above for the uniqueness of $F_{ab}$.  If  $(\overset{1}{J}\,^{a} - \overset{2}{J}\,^{a})  \xi_a = 0$ for all  four-velocity vectors  $\xi^a $ at all points, then $\overset{1}{J}\,^{a} =  \overset{2}{J}\,^{a}$.} So in this case, schematically, we have:

 \[
\left.
\begin{array} {lc}
{}\hspace{4em}&\textrm{invariant map from}  \\
{}&\textrm{tangent lines to charge density values}   \\
{}& +  \\
{}&\textrm{temporal orientation}  
\end {array}
 \right\}  \hspace{1em} 
 \Longrightarrow \hspace{1em}    J^a 
 \]
 \vspace{0em}

With the fields $(F_{ab}, J^a)$ in hand,  we can formulate the most basic laws of classical electromagnetic theory:  (i) the (Lorentz) equation of motion for a charged test particle in the presence of an electromagnetic field, and (ii)  Maxwell's field equations.  

Once again, we can represent the world line of a test particle as the image of a smooth timelike curve of form  $\gamma\!:I \rightarrow M$, where $I$ is a connected subset of $\mathbb{R}$. For convenience, we assume that the curve is so  parametrized that the tangent vector $\xi^a$ to the curve at every point  is of unit length and future-directed. Then  $\xi^a$ is called the \emph{four-velocity} of the particle at the point, and the directional derivative $\xi^b \, \nabla_b \, \xi^a$ there is called its \emph{four-acceleration}.  All the ingredients are now assembled to insert into
``$F = ma$". Suppose the particle has mass $m \ne 0$ and charge $q$. Then at every point, the force acting on it is $q \, F^{a}_ {\ b} \, \xi^b$.  So its equation of motion comes out as:
\begin{equation}  \label{Lorentz.equation.motion}
q \, F^{a}_ {\ b} \, \xi^b \, = \,  m \ \xi^b \, \nabla_b \, \xi^a.
\end{equation}
%Note that the equation makes geometric sense since the force vector on the left is  orthogonal to $\xi^a$.   (The four-acceleration $\xi^b \, \nabla_b \, \xi^a$  is automatically orthogonal to $\xi^a$.\footnote{This  follows since $(\xi^a \,\xi_a) = 1$, and therefore  $(\xi^b \, \nabla_b \, \xi^a) \, \xi_a =  \frac{1}{2}  \xi^b \, \nabla_b \, (\xi^a \,\xi_a) = \frac{1}{2}  \xi^b \, \nabla_b \, (1)  = 0.$} So the vector on the  left had better be orthogonal to it as well.) 
Note that the particle's four-acceleration at every point is independent of the background temporal orientation. (If we reparametrize $\gamma$ so that the tangent vector to the curve at every point is of unit length and \emph{past-directed}, i.e., so that it is $-\xi^a$ rather than $\xi^a$, then its four-acceleration is $(-\xi^b) \, \nabla_b \, (-\xi^a) = \xi^b \, \nabla_b \, \xi^a$.) 
%I will return to this point in the next section.  

In our notation, Maxwell's equations come out as:
 \begin{eqnarray}
\nabla_{[a} \, F_{bc]} & = & \mathbf{0}   \label{first.invariant.Maxwell} \\
\nabla_n \, F^{na} & = & J^a.   \label{second.invariant.Maxwell}  
\end{eqnarray}
It is not my purpose to develop electromagnetic theory, but it is worth mentioning two immediate consequences of these equations.  (i) The first asserts that $F_{ab}$ is closed. So, by the converse to the Poincar\'{e} lemma, it is, at least locally, exact, i.e.,  given any point $p$ in $M$, there is an open set $O$ containing $p$, and a  smooth field $A_a$ on $O$, such that $F_{ab} = \nabla_{[a}  A_{b \, ]}$. (ii) The second equation implies that $\nabla_a J^a = 0$, which captures the requirement of local charge conservation.\footnote{Since $F^{ab}$ is anti-symmetric, we have
$\nabla_a J^a = \nabla_a \nabla_n \, F^{na} = \nabla_{[a} \nabla_{n]} \, F^{na}$.  But if $R^a_{\ bcd}$  is the Riemann curvature field associated with $\nabla_n$, and $R_{bc} = R^a_{\ bca}$ is the associated Ricci curvature field, we also have
\[
\begin{array}{lllll}
2 \, \nabla_{[a} \nabla_{n]} \, F^{na} & =&   -R^n_{\ man} \, F^{ma} - R^a_{\ man} \, F^{nm} 
&  = &    -R^n_{\ man} \, F^{ma} + R^a_{\ mna} \, F^{nm}  \\
{}   & =&   -R_{ma} \, F^{ma} + R_{mn} \, F^{nm}  
& =&    -R_{mn} \, F^{mn} + R_{nm} \, F^{mn}   \\
{}   & = &  -R_{mn} \, F^{mn} + R_{mn} \, F^{mn} & =& \mathbf{0}.  
\end{array}
\]
All the component assertions here follow from basic facts about $R^a_{\ bcd}$ and $R_{bc}$. (See Wald (1984), section 3.2.  Warning:  the sign conventions in that book differ slightly from mine.) The first equality holds for any tensor field $F^{ab}$; anti-symmetry plays no role.  The second follows from the fact that $R^a_{\ b(cd)} = \mathbf{0}$. The third and fourth, respectively, involve nothing more than the definition of $R_{ab}$ and systematic index substitution. The fifth equality follows from the symmetry of $R_{mn}$.}

\section{Time Reversal Invariance}

As already indicated, the objects $F_{ab}$ and $J^a$ used to represent the electromagnetic field and its charge-current source field are only defined relative to a background choice of temporal orientation.  The time reversal operation is naturally understood as one taking fields on $M$ as determined relative to one temporal orientation to corresponding fields on $M$  as determined relative to the other.
%\footnote{This characterization of the time reversal operation is, on the surface, different from the standard one discussed in section 2.  I will argue in a moment that it captures the very same idea and is better suited to the context of relativity theory.} 
In the case at hand, we thus have an operation of form 
\begin{center}
$(M, \, g_{ab}, \,  \tau^a,  \,  F_{ab}, \, J^a)  \hspace{2em}  \stackrel{time \  reversal}{\longmapsto}
\hspace{2em}  (M, \, g_{ab}, \,  -\tau^a,  \, {}^T\hspace{-.2em}F_{ab}, \,  {}^T\hspace{-.2em}J^a)$
\end{center}
to consider.\footnote{I hope my imperfect notation does not cause confusion.  I am using the same symbol `\emph{T}' for the time reversal operator that I used in section 2, even though the operator is understood somewhat differently here. In a sense that can be made precise, the two ways of understanding it are equivalent if the background relativistic spacetime  admits a ``frame of reference".  (See section 6.) In particular, they are equivalent in Minkowski spacetime.}   Here ${}^T\hspace{-.2em}F_{ab}$ and ${}^T\hspace{-.2em}J^a$ are understood to be the  unique tensor fields on $M$ that  bear the same relation to $-\tau^a$ that $F_{ab}$ and $J^a$  bear to $\tau^a$.

I claim first that
\begin{eqnarray}
 {}^T\hspace{-.2em}F_{ab}  &=&  -F_{ab}   \label {T-F}   \\
 {}^T\hspace{-.2em}J^a  &=& - J^a. \label {T-J}
 \end{eqnarray}
The arguments for the two cases are very much the same. Recall,  first, our route from the force map $<\hspace{-.4em}L, \, q\hspace{-.4em}> \hspace{.2em} \mapsto F(L, q)$ to the tensor field $F_{ab}$. At a crucial stage we traded in tangent lines for unit vectors. Here is a complete statement of the condition that characterizes $F_{ab}$: 
for all points $p$,  all charge values $q$, and all timelike lines $L$ in the tangent space at $p$, the force $F(L, q)$ on a test particle at $p$ with charge $q$ and  (invariant) instantaneous velocity $L$ is given by
\begin{equation} \label{first force equation}
F(L, q)  =  q \  F^{a}_ {\ b} \, \xi^b,
\end{equation}   
where $\xi^a$ is the unit timelike vector at $p$, co-alligned with $L$,  \emph{that is future directed with respect to $\tau^a$}.  We need  to consider what happens if we keep this characterization intact, but replace $\tau^a$ in the italicized expression by $-\tau^a$, i.e., if we represent the tangent line $L$, not by the unit vector $\xi^a$, but rather by $-\xi^a$.\footnote{I take for granted that the charge (and mass) of a particle are not dependent on temporal orientation. One could  challenge this assumption, perhaps, but Albert shows no inclination to do so. His  concerns are entirely different, and cannot even be reached unless this much is assumed.}  In that case we are led to an anti-symmetric tensor field ${}^TF_{ab}$ on $M$ that satisfies an equation parallel in form to (\ref{first force equation}), namely
\begin{equation} \label{second force equation}
F(L, q)  =  q \  ({}^TF^{a}_ {\ b}) \, (-\xi^b).
\end{equation}
To determine ${}^TF_{ab}$, we need only compare (\ref{second force equation}) with (\ref{first force equation}). It follows from the two conditions that, at all points, 
$({}^TF^{a}_ {\ b} \,  +  F^{a}_ {\ b})  \, \xi^b = \mathbf{0}$,  for all unit timelike vectors $\xi^a$ future-directed with respect to $\tau^a$.\footnote{Of course, we could just as well formulate this as the  condition that  $({}^TF^{a}_ {\ b} \,  +  F^{a}_ {\ b})  \, \eta^b = \mathbf{0}$  for all unit timelike vectors $\eta^a$ future-directed with respect to $-\tau^a$.} This  implies that $ {}^T\hspace{-.2em}F_{ab} = -F_{ab}$ \, (since the set of all such vectors at any point spans the tangent space there). 

Consider next the corresponding argument for (\ref{T-J}). The complete statement of the condition characterizing $J^a$ is this:   for all points $p$, and all timelike lines $L$ in the tangent space at $p$, the charge density determined by an observer at $p$ with (invariant) instantaneous velocity $L$ is given by
\begin{equation} \label{first charge equation}
J(L)  =  J^{a}\, \xi_a,
\end{equation}  
where $\xi^a$ is the unit timelike vector at $p$, co-alligned with $L$,  \emph{that is future directed with respect to $\tau^a$}.   If we replace $\tau^a$ by $-\tau^a$ in the italicized expression, we are led to a field ${}^TJ^a$ on $M$ satisfying an equation parallel in form to (\ref{first charge equation}), namely
\begin{equation} \label{second charge equation}
J(L)  =  ({}^TJ^{a}) \, (-\xi_a).\footnote{Here I am assuming that charge density at a point, as determined by an observer, does not depend on temporal orientation.}
\end{equation} 
It then follows from (\ref{first charge equation}) and (\ref{second charge equation}) that $({}^TJ^{a} \, +  J^{a})  \, \xi_a = \mathbf{0}$,  for all unit timelike vectors $\xi^a$ future-directed with respect to $\tau^a$.  And this, in turn, implies that
${}^T\hspace{-.2em}J^a  = - J^a$. 
   
Suppose now we agree that  ${}^T\hspace{-.2em}F_{ab}  =  -F_{ab}$ and ${}^T\hspace{-.2em}J^a  =  -J^a$. Then it follows immediately that  Maxwell's equations are time reversal invariant. (That is, the initial fields $(F_{ab}, \, J^a)$ satisfy the equations iff the time reversed fields $({}^T\hspace{-.2em}F_{ab}, \,  {}^T\hspace{-.2em}J^a)$ do.) 
\[
\begin{array} {lll}
\nabla_{[a} \, F_{bc]} = \mathbf{0}    &\Longleftrightarrow  \  \   \nabla_{[a} \,(-F_{bc]}) = \mathbf{0}    \  &\Longleftrightarrow  \  \     \nabla_{[a} \,({}^T\hspace{-.2em}F_{bc]})  = \mathbf{0}    \\
\nabla_n \, F^{na}  =  J^a    &\Longleftrightarrow  \  \    \nabla_n \, (-F^{na})  = - J^a    &\Longleftrightarrow   \ \   \nabla_n \, ({}^T\hspace{-.2em}F^{na})  = {}^T\hspace{-.2em}J^a.     
\end{array}
\]
The computations here are trivial. But it is an important, non-trivial property of Maxwell's equations that they \emph{are} time reversal invariant. This outcome was not guaranteed by our characterization of $F_{ab}$,  $J^a$, and the time reversal operator T.  Consider, for example, the following alternative (totally unmotivated) field equation:
\begin{equation} \label{adhoc equation}
\nabla_a (F_{mn} \, F^{mn}) = J_a. 
\end{equation}
Clearly, this one is \emph{not} time reversal invariant. The left side tensor field is invariant under time reversal:  $\nabla_a \left( ({}^TF_{mn}) \, ({}^T F^{mn}) \right) = \nabla_a (F_{mn} \, F^{mn})$.   But the right side field is not:   $(^TJ_a) = -J_a \ne J_a$.  The difference between (\ref{adhoc equation}) and the second Maxwell equation ($\nabla_n \, F^{na}  =  J^a$) just comes to this. In both cases, the tensor fields asserted to be equal depend on (or are defined relative to) the background temporal orientation. But only in the second case do the dependencies match.    

\section{Decomposition of the Electromagnetic Field}

I have argued that classical electromagnetic theory is time reversal invariant without making any reference to component electric and magnetic fields whatsoever.  I turn to the latter in this section. 

So far, we have under consideration a relativistic spacetime $(M, g_{ab})$, a continuous timelike vector field $\tau^a$ on $M$, and a pair of smooth fields $(F_{ab}, J^a)$ representing the electromagnetic field and its charge-current source field (relative to the temporal orientation determined by $\tau^a$).  Further background geometric structure must be introduced if we are to decompose $F_{ab}$ and $J^a$. (It is only relative to that additional structure that the decomposition is defined.)

A \emph{volume element} on $(M, g_{ab})$ is a smooth tensor field $\epsilon_{abcd}$ on $M$ that is completely anti-symmetric ($\epsilon_{[abcd]} = \epsilon_{abcd}$) and satisfies the normalization condition  $\epsilon_{abcd} \, \epsilon^{abcd} = -24$.\footnote{Given an ordered set  of vectors $(\overset{1}{\xi} \,^{a}, \, \overset{2}{\xi} \,^{a},  \, \overset{3}{\xi} \,^a,  \, \overset{4}{\xi} \,^a)$ at a point, we can think of $\epsilon_{abcd} \, \overset{1}{\xi}\,^{a} \, \overset{2}{\xi} \,^{b}  \, \overset{3}{\xi} \,^c  \, \overset{4}{\xi} \,^d $ as the ``oriented volume" of the parallelepiped determined by the vectors. Orientation enters in that the order of the vectors in the quadruple matters, e.g., $\epsilon_{abcd} \, \overset{2}{\xi}\,^{a} \, \overset{1}{\xi} \,^{b}  \, \overset{3}{\xi} \,^c  \, \overset{4}{\xi} \,^d = - \epsilon_{abcd} \, \overset{1}{\xi}\,^{a} \, \overset{2}{\xi} \,^{b}  \, \overset{3}{\xi} \,^c  \, \overset{4}{\xi} \,^d.$ The normalization condition $\epsilon_{abcd} \, \epsilon^{abcd} = -24$ is equivalent to the requirement that, for all orthonormal bases $\{ \overset{1}{\xi} \,^{a}, \, \overset{2}{\xi} \,^{a},  \, \overset{3}{\xi} \,^a,  \, \overset{4}{\xi} \,^a \}$ at all points, $\epsilon_{abcd} \, \overset{1}{\xi}\,^{a} \, \overset{2}{\xi} \,^{b}  \, \overset{3}{\xi} \,^c  \, \overset{4}{\xi} \,^d = \pm 1$. }  If there exists a volume element on $(M, g_{ab})$, it is said to be \emph{orientable}. In this case, there exist exactly two volume elements on $(M, g_{ab})$ altogether.  If $\epsilon_{abcd}$ is one, the other is $-\epsilon_{abcd}$. 

In what follows, I will assume that our background spacetime $(M, g_{ab})$ is orientable (as well as temporally orientable), and take  $\epsilon_{abcd}$ to be a volume element on $(M, g_{ab})$.  

 I  will  also assume that $(M, g_{ab})$ admits what I will call a ``frame of reference", namely  a smooth, future-directed, unit  timelike vector field $\eta^a$ on $M$ that  is constant (i.e., $\nabla_a  \, \eta^b = \mathbf{0}$).\footnote{The condition of constancy is more stringent than necessary here, but it will allow us to sidestep  certain complexities that are of no special importance for our purposes. Indeed, it would be no great loss if we restricted attention in this section to Minkowski spacetime (which certainly admits ``frames of reference"). } Let $\eta^a$ be one such in what follows. It determines, at least locally, a foliation of spacetime into a family $\mathcal{S}$ of  spacelike hypersurfaces  that are everywhere orthogonal to $\eta^a$.\footnote{If $\eta^a$ is constant, it is closed ($\nabla_{[a} \, \eta_{b]} = \mathbf{0}$). So, at least locally, it is exact, i.e., there exists a smooth scalar field $t$ such that $\eta_a = \nabla_a \, t$.  The $t$ = constant hypersurfaces are spacelike, and everywhere orthogonal to $\eta^a$.} These hypersurfaces will play the role of the simultaneity slices taken for granted in section 2. 
 
 We can now recover the standard textbook formulation of classical electromagnetic theory by decomposing $F_{ab}$ and $J^a$ relative to $\eta$ and $\epsilon_{abcd}$. Consider the following fields on $M$:
\begin{eqnarray}
h^{a}_ {\ b}  & = &  g^{a}_ {\ b}  - \eta^a  \, \eta_b  \label{def.h} \\
\rho & = & J^a \, \eta_a  \label{def.mu}\\
j^a & = &  h^{a}_ {\ b} \, J^b  \label{def.j}\\    
E^a & = & F^{a}_ {\ b} \, \eta^b  \label{def.E}\\
B^a & = & \frac{1}{2} \, \epsilon^{abcd} \, \eta_b \, F_{cd}.   \label{def.B}
\end{eqnarray}
\indent Let's consider them in turn. We can think of $h^{a}_ {\ b}$, first, as a projection field that, at every point, projects vectors onto their components orthogonal to $\eta^a$. To see this, consider a vector $\lambda^a$ at some point. We can express it in the form: $\lambda^a = (\lambda^b  \eta_b) \, \eta^a + (g^{a}_ {\ b}  - \eta^a   \eta_b) \, \lambda^b$. The first term on the right is proportional to $\eta^a$. The second is orthogonal to $\eta^a$ (since $\eta^a \eta_a = 1$ and, therefore, $((g^{a}_ {\ b}  - \eta^a   \eta_b) \, \lambda^b) \eta_a = (\lambda^a  \eta_a) - (\eta_b \lambda^b) = 0$).  In particular (taking $J^a$ for $\lambda^a$), we have $J^a = (J^b  \eta_b) \, \eta^a + h^{a}_ {\ b}\, J^b$, i.e.,  
\begin{equation}
J^a = \rho  \, \eta^a +  j^a.
\end{equation}
\noindent{$\rho$ is the  \emph{charge-density relative to $\eta^a$}, and \emph{$j^a$ is the current-density relative to $\eta^a$}. (We have already encountered the former in our characterization of $J^a$.)  Since $j^a$ is orthogonal to $\eta^a$ at every point, it is everywhere tangent to the hypersurfaces in $\mathcal{S}$. So we can think of $j^a$ as ``residing" on them.  (This will be true of $E^a$ and $B^a$ as well.) Note that $\nabla_a \,  h^{b}_ {\ c} = 0$, since $\nabla_a \,  g_ {bc}  = 0 = \nabla_a  \, \eta^{b}$.  

 Next, $E^a$ is the \emph{electric field} as determined relative to $\eta^a$. It is orthogonal to $\eta^a$ at every point (since $F_{ab}$ is anti-symmetric). It follows from our characterization of $F_{ab}$ that $E^a$ is also the undifferentiated, net electromagnetic force that would be felt by a test particle of unit positive charge with four-velocity $\eta^a$. This can be captured as the standard claim that, as determined relative to any particular frame of reference, a test particle at rest in the frame experiences an electric force, but no magnetic force.
 
$B^a$ is the \emph{magnetic field} as determined relative to $\eta^a$  \emph{and} $\epsilon_{abcd}$. The latter dependency is crucial.  It too is orthogonal to $\eta^a$ at every point (since $\epsilon^{abcd}$ is completely anti-symmetric and so $\epsilon^{abcd} \eta_a \eta_b = 0$). Thus, $E^a$ and $B^a$ are tangent everywhere to the hypersurfaces in  $\mathcal{S}$.  

It follows from (\ref{def.h}) --  (\ref{def.B}), with just a bit of computation, that:
\begin{equation}
F_ {ab}  =  -2  \, \eta_{[a} \,E_  {b]} +   \epsilon_{abcd}  \, \eta^c  B^d.  \label{Fdecomp} \\
\end{equation}
Again, it is not my purpose to develop electromagnetic theory, but it is worth mentioning two consequences of (\ref{Fdecomp}):
\[
\begin{array}{lll}
F^{ab} \, F_{ab}  &=& 2 \, (E^a  E_a - B^a  B_a)     \\
\epsilon_{abcd} F^{ab} F^{cd} & = & 8 \  E^a B_a.    
\end{array}
\]

\noindent They establish that $(E^a  E_a - B^a  B_a)$ and $E^a B_a$ are invariant, i.e., have the same value for all choices of the background frame of reference $\eta^a$. (In three-vector notation, they would be expressed as  $(|\textbf{B}|^2 - |\textbf{E}|^2)$ and $-\textbf{E} \cdot \textbf{B}$.\footnote{The sign changes arise because $E^a$ and $B^a$ are spacelike.}) It also follows from (\ref{Fdecomp}) and (\ref{def.h}) --  (\ref{def.B}) that we have the following equivalences:

\[
\begin{array} {lll}
\nabla_{[a} \, F_{bc]}  =  0   & \Longleftrightarrow &   
\left\{
\begin{array}{l} 
\nabla_a \, B^a = 0   \\
\epsilon^{anbc} \,  \eta_n  \nabla_b \,  E_c  = -\eta^n \,  \nabla_n  \, B^a 
\end{array} 
\right.  \\
{} & {} & {} \\
\nabla_n \, F^{na} =  J^a  &  \Longleftrightarrow  &
\left\{
\begin{array} {l}
\nabla_a \, E^a = \rho  \\
\epsilon^{anbc} \,  \eta_n  \nabla_b \,  B_c  = \eta^n \,  \nabla_n  \, E^a + j^a. 
\end{array} 
\right.
\end{array}
\] 

\noindent Each of the  equations on the right captures, in our notation, one of the four Maxwell equations, resp.  (\ref{3rd.3dim.Maxwell}), (\ref{2nd.3dim.Maxwell}), (\ref{1st.3dim.Maxwell}), (\ref{4th.3dim.Maxwell}).

\section{Time Reversal Invariance Once Again}

We need to determine how the time reversal operator $T$ acts on $\rho, j^a, E^a$, and $B^a$.  It will help to first consider, side by side,  three senses of orientation:  temporal, spatial, and spatiotemporal.  

The first and third are determined directly by $\tau^a$ and $\epsilon_{abcd}$ respectively.  We have encountered them already. A timelike vector $\xi^a$ at a point is \emph{future directed} or \emph{past directed}  relative to $\tau^a$, of course, depending on whether $\tau^a \xi_a$ is positive or negative. Similarly, an ordered  set of linearly independent vectors  $(\overset{1}{\xi} \,^{a}, \, \overset{2}{\xi} \,^{a},  \, \overset{3}{\xi} \,^a,  \, \overset{4}{\xi} \,^a)$ at a point has \emph{positive spatiotemporal orientation} or \emph{negative spatiotemporal orientation}  relative to $\epsilon_{abcd}$ depending on whether its oriented volume $\epsilon_{abcd} \, \overset{1}{\xi}\,^{a} \, \overset{2}{\xi} \,^{b}  \, \overset{3}{\xi} \,^c  \, \overset{4}{\xi} \,^d$ is positive or negative.\footnote{It cannot be $0$ since $\epsilon_{abcd} \, \overset{1}{\xi}\,^{a} \, \overset{2}{\xi} \,^{b}  \, \overset{3}{\xi} \,^c  \, \overset{4}{\xi} \,^d = 0$ iff the set $\{\overset{1}{\xi} \,^{a}, \, \overset{2}{\xi} \,^{a},  \, \overset{3}{\xi} \,^a,  \, \overset{4}{\xi} \,^a\}$ is linearly dependent.} The notion of spatial orientation is slightly more delicate, because it presupposes a foliation of spacetime into a family of spacelike hypersurfaces (that represent ``space" at a given ``time").  

Once again, let $\eta^a$ be a frame of reference on $(M, g_{ab})$, and let $\mathcal{S}$ be the associated family of spacelike hypersurfaces everywhere orthogonal to $\eta^a$. We may take a \emph{spatial volume element} relative to $\eta^a$ to be a smooth field $\epsilon_{abc}$ that is completely anti-symmetric ($\epsilon_{[abc]} = \epsilon_{abc}$), normalized so that $\epsilon_{abc} \, \epsilon^{abc} = -6$, and orthogonal to $\eta^a$, i.e., $\epsilon_{abc} \eta^a = \mathbf{0}$.\footnote{Clearly, it follows from anti-symmetry that if  $\epsilon_{abc} \eta^a = \mathbf{0}$, then also $\epsilon_{abc} \eta^b = \epsilon_{abc} \eta^c = \mathbf{0}$. The normalization condition $\epsilon_{abc} \, \epsilon^{abc} = -6$ is equivalent to the requirement that, for all orthonormal bases of form $\{\eta^a, \,  \overset{1}{\sigma} \,^{a}, \, \overset{2}{\sigma} \,^{a},  \, \overset{3}{\sigma} \,^a\}$, at all points, $\epsilon_{abc} \, \overset{1}{\sigma}\,^{a} \, \overset{2}{\sigma} \,^{b}  \, \overset{3}{\sigma} \,^c  = \pm 1$.} We can think of $\epsilon_{abc}$  as residing on the hypersurfaces in $\mathcal{S}$.  An ordered  set of linearly independent vectors  $(\overset{1}{\sigma} \,^{a}, \, \overset{2}{\sigma} \,^{a},  \, \overset{3}{\sigma} \,^a)$ at a point, all orthogonal to $\eta^a$, has \emph{positive spatial orientation} or \emph{negative spatial orientation}  relative to $\epsilon_{abc}$ depending on whether $\epsilon_{abc} \, \overset{1}{\sigma}\,^{a} \, \overset{2}{\sigma} \,^{b}  \, \overset{3}{\sigma} \,^c$ is positive or negative. There are exactly two spatial volume elements relative to $\eta^a$ on  $(M,g_{ab})$, namely  $\epsilon_{anbc} \eta^n$ and  $-\epsilon_{anbc} \eta^n$.\footnote{The choice for the index of contraction does not matter. I use the second index to match the choice in section 3.}

I have construed time reversal as an operation taking fields on $M$ as determined relative to $\tau^a$ to corresponding fields as determined relative to $-\tau^a$.  We can understand ``spatiotemporal parity reversal" and  ``spatial parity reversal" (with ``space" taken relative to $\eta^a$)  similarly. Specifically, we can take the former to be an operation taking fields as determined relative to $\epsilon_{abcd}$ to fields as determined relative to $-\epsilon_{abcd}$. And we can take the latter to be one taking fields as determined relative to $\epsilon_{abc} =\epsilon_{anbc} \eta^n$ to ones as  determined relative to $-\epsilon_{abc} = -\epsilon_{anbc} \eta^n$.

The three operations under consideration are not independent of each other. The time reversal operation takes $\eta^a$ to $-\eta^a$. So, for example, a combination of time reversal and spatiotemporal parity reversal ($\epsilon_{abcd} \mapsto -\epsilon_{abcd})$ leaves spatial orientation intact:
$\epsilon_{abc} = \epsilon_{anbc} \eta^n  \mapsto (-\epsilon_{anbc}) (-\eta^n) = \epsilon_{abc}$.
There are three composite possibilities in all.

\begin{equation*}
\begin{array}{llll} 
\textrm{\textbf{T} \ time reversal without spatial parity reversal:}  &  \tau^a  & \mapsto  &    -\tau^a    \\ 
{} & \epsilon_{abcd} &\mapsto & -\epsilon_{abcd} \\
\vspace{2em}  
{} & \epsilon_{abc} &\mapsto & \epsilon_{abc} \\

\textrm{\textbf{P} \ spatial parity reversal without time reversal:}  &  \tau^a  & \mapsto  &    \tau^a \\   
{} & \epsilon_{abcd} &\mapsto & -\epsilon_{abcd} \\
\vspace{2em}  
{} & \epsilon_{abc} &\mapsto & -\epsilon_{abc} \\

\textrm{\textbf{TP} \ time reversal and spatial parity reversal:}  &  \tau^a  & \mapsto  &    -\tau^a  \\
{} & \epsilon_{abcd} &\mapsto & \epsilon_{abcd} \\
\vspace{2em}  
{} & \epsilon_{abc} &\mapsto & -\epsilon_{abc} 
\end{array}
\end{equation*}

Let's consider how all three operations $\mathbf{T}$, $\mathbf{P}$, and $\mathbf{TP}$ act on $F_{ab}$, $J^a$,  and the fields that figure in their decomposition. The full set comes out as in Table 1.
\vspace{1em}

\begin{equation*}
\begin{array}{lllllllll} 
{}^TF_{ab} & = & -F_{ab}  \hspace{2.5em}   &{}^PF_{ab} &=& F_{ab} \hspace{2.5em} &{}^{TP}F_{ab}  & = & -F_{ab}\\ 
{}^TJ^a & = & -J^a   \hspace{2em}   &{}^PJ^a &=& J^a  \hspace{2em}&{}^{TP}J^a  & = & -J^a\\
\vspace{-.5em} 
{} & {} & {}     &{} &{}& {} &{} & {} & {}\\ 
{}^T\eta^a  & = & -\eta^a   \hspace{2em}   &{}^P\eta^a &=& \eta^a \hspace{2em} &{}^{TP}\eta^a  & = & -\eta^a \\
{}^Th^a_{\ b} & = & h^a_{\ b}  \hspace{3em}   &{}^Ph^a_{\ b} & = & h^a_{\ b}  \hspace{3em} & {}^{TP}h^a_{\ b} & = & h^a_{\ b} \\
\vspace{-.5em} 
{} & {} & {}     &{} &{}& {} &{} & {} & {}\\ 
{}^T\rho & = & \rho  \hspace{2em}   & {}^P\rho & = & \rho  \hspace{2em}&{}^{TP}\rho & = & \rho  \\
{}^Tj^a & = &  -j^a   \hspace{2em}   & {}^Pj^a &=& j^a  \hspace{2em}& {}^{TP}j^a  & = & -j^a \\
{}^TE^a & = & E^a   \hspace{2em}   & {}^PE^a  &=& E^a  \hspace{2em}& {}^{TP}E^a  & = & E^a \\
{}^TB^a & = & -B^a   \hspace{2em}   & {}^PB^a  &=& -B^a  \hspace{2em}& {}^{TP}B^a  & = & B^a
\end{array}
\vspace{1em}
\end{equation*}
\begin{table}[h]
  %\centering 
  \begin{center}
\setlength{\unitlength}{1cm}
\begin{picture}(0,0) 
%\put(0.55,2.25) {\framebox(3.7,3.1)}
%\put(9.45, 2.25) {\framebox(3.8,3.1)}
\put(1.3,2.) {\framebox(3.3,2.9)}
\put(9.15, 2.) {\framebox(3.3,2.9)}
  %\put(0.0,0.0){\epsfig{figure=psfiles/Figure3.eps}}
  %\put(1.55,2.92){electromagnetic force vector}
  %\put(-2.9, 3.5){\normalsize four-velocity?  $\xi^a$}
\end{picture} 
 \begin{minipage}[b]{14.0cm}

\vspace{-1.5em}
  \caption{ The action of the operations $\mathbf{T}$, $\mathbf{P}$, and $\mathbf{TP}$  on $F_{ab}$, $J^a$,  and the fields that figure in their decomposition. The left box shows  the standard time reversal transformations. The right box shows the ``Albert alternative transformations".}\label{table }
\end{minipage}
\end{center}
\end{table}

\noindent The entries in the third column are determined by the those in the first two, since $\mathbf{TP}$ is the product, or composition, of $\mathbf{T}$ and $\mathbf{P}$  (in either order).\footnote{So, for example, ${}^{TP}B^a = {}^{T} ({}^{P}B^a) = {}^{T} (-B^a) = {}^{T} ({}^{T}B^a) = B^a.$  (The final step follows from the modest assumption that the time reversal operation is self-cancelling in the sense that $\mathbf{T}^2 = \mathbf{I} $.)}  The entries in the second column are all straightforward except for the final one.  None of the listed fields other than
$B^a  =  \frac{1}{2} \, \epsilon^{abcd} \, \eta_b \, F_{cd} = \frac{1}{2} \, \epsilon^{acd} \, F_{cd} $ depends on $\epsilon_{abc}$; so none is affected by the operation of parity reversal.  For $B^a$ itself, we have
\[
{}^PB^a  =  \frac{1}{2} \, ({}^P\epsilon^{acd}) \, ({}^PF_{cd}) = \frac{1}{2} \, (-\epsilon^{acd}) \, F_{cd} = -B^a.\footnote{Note that $\mathbf{T}$ and $\mathbf{P}$ (and, so, also $\mathbf{TP}$) commute with the operations of raising and lowering indices. This follows since neither $g_{ab}$ nor its inverse $g^{ab}$  depends on the background temporal or spatial orientation, and therefore ${}^Tg_{ab} = g_{ab} = {}^Pg_{ab}$ and ${}^Tg^{ab} = g^{ab} = {}^Pg^{ab}$. So, for example,
\[
{}^P\epsilon^{acd} = {}^P(g^{am} \, g^{cn} \, g^{dp} \,\epsilon_{mnp}) = ({}^Pg^{am}) \, ({}^Pg^{cn}) \, ({}^Pg^{dp}) \, ({}^P\epsilon_{mnp}) =  g^{am} \, g^{cn} \, g^{dp} \, (-\epsilon_{mnp}) = -\epsilon^{acd}.  
\]}      
\]
That leaves only the entries in the first column to check.  We have already argued that ${}^TF_{ab}  = -F_{ab}$ (\ref{T-F}) and ${}^TJ^a  =  -J^a$ (\ref{T-J}).  The time reversal operation flips future-directed timelike vectors to past-directed ones. So we have ${}^T\eta^a  =  -\eta^a$ and, therefore,
${}^Th^a_{\ b}  =  {}^T(g^a_{\ b}  - \eta^a \, \eta_b)  = [({}^Tg^a_{\ b}) -  ({}^T\eta^a)  ({}^T\eta_b)] = [g^a_{\ b} -  (-\eta^a)  (-\eta_b)] = h^a_{\ b}.$  The final four entries in the first column are  easy to check as well. 
\begin{equation*}
\begin{array}{lllllllll} 
{}^T\rho & = & ({}^TJ^a) \, ({}^T\eta_a) &=& (-J^a) (-\eta_a) &=& \rho      \\
{}^Tj^a & = &   ({}^Th^{a}_ {\ b})   \, ({}^T J^b) &=&  h^{a}_ {\ b}  \, (-J^b) &=& -j^a   \\    
{}^TE^a & = & ({}^TF^{a}_ {\ b}) \, ({}^T\eta^b) &=& (-F^{a}_ {\ b}) \, (-\eta^b) &=& E^a \\ 
{}^TB^a  &=&  \frac{1}{2} \, ({}^T\epsilon^{acd}) \, ({}^TF_{cd}) &=& \frac{1}{2} \, \epsilon^{acd} \, (-F_{cd}) &=& -B^a.  
\end{array}
\end{equation*}

I have bothered to produce the table because it helps to clarify the relation of the standard time reversal transformations (\ref{STRT1}) - (\ref{STRT4}) to David Albert's proposed alternatives (\ref{ATRT1}) - (\ref{ATRT4}). They correspond, respectively, to the final four entries in the $\mathbf{T}$ and  $\mathbf{TP}$ columns. This really is the heart of the matter.  \emph{Albert's transformations should properly be seen as characterizing a composite operation of time \emph{and} parity reversal.} Magnetic fields do not ``just lie there" (in the sense discussed in section 2 under time reversal.   But they \emph{are} left intact under $\mathbf{TP}$ because both component operations invert them (and the two actions cancel each other).

I should anticipate one possible point of confusion. It was asserted in section 2 that Maxwell's equations are not (all) invariant under Albert's alternative transformations. Now it is claimed that those transformations characterize the $\mathbf{TP}$ operation. Doesn't if follow that Maxwell's equations are \emph{not} invariant under the composite operation of time and parity reversal? (And, if so, doesn't that conclusion fly in the face of orthodoxy just as much as Albert's claim that they are not invariant under time reversal alone?) The answer to the first question is  `no'.  Here is why in the language of section 2. Consider, for example, equation (\ref{2nd.3dim.Maxwell}):  $\nabla   \times \textbf{E}   =   - \frac{ \partial \textbf{B}}{ \partial t}$. It is \emph{not} invariant under Albert's transformations if they are understood as time reversal transformations. For in that case, $\textbf{E}$ and $\textbf{B}$ are left intact, and the partial derivative operator   $\frac{ \partial {}} { \partial t}$  is replaced by $\frac{ \partial {}} { \partial(- t)} = -\frac{ \partial {}} { \partial t}$. But (\ref{2nd.3dim.Maxwell}) \emph{is} invariant under Albert's transformations if they are understood as combined time and parity reversal transformations. In the latter case, cross-products and curls (such as $\nabla   \times \textbf{E}$) are also multiplied by $-1$, and that restores the balance between the two sides of the equation.} 

The transformation properties for $B^a$ are exactly the same as for angular velocity.
% (the canonical example of an ``axial" vector field). 
Let $\xi^a$ be a (not necessarily constant)  future-directed, unit timelike vector field on $M$.  We may think of  $\xi^a$ as the four-velocity field of a fluid.  The \emph{angular velocity} field of the fluid is then given by:
\[
\omega^a  =  \frac{1}{2} \, \epsilon^{abcd} \, \xi_b \, \nabla_c  \, \xi_d.
\]
Applying the $\mathbf{T}$, $\mathbf{P}$, and $\mathbf{TP}$ to $\omega^a$ yields:
\begin{equation*}
\begin{array}{lllllll}
 {}^T\omega^a  &=& \frac{1}{2} \,  ({}^T\epsilon^{abcd}) \,  ({}^T\xi_b) \,  \nabla_c  \,  ({}^T\xi_d) &=&
 \frac{1}{2} \,  (-\epsilon^{abcd}) \,  (-\xi_b)  \, \nabla_c  \,  (-\xi_d) &=& -\omega^a \\
   {}^P\omega^a  &=& \frac{1}{2} \,  ({}^P\epsilon^{abcd}) \,  ({}^P\xi_b) \,   \nabla_c  \,  ({}^P\xi_d) &=&
 \frac{1}{2} \,  (-\epsilon^{abcd}) \,  \xi_b  \,  \nabla_c  \,  \xi_d  &=& -\omega^a \\
 {}^{TP}\omega^a  &=& \frac{1}{2} \,  ({}^{TP}\epsilon^{abcd}) \,  ({}^{TP}\xi_b)   \, \nabla_c  \,  ({}^{TP}\xi_d) &=&
 \frac{1}{2} \,  \epsilon^{abcd} \,  (-\xi_b)  \, \nabla_c  \,  (-\xi_d) &=& \omega^a. 
 \end{array}
 \end{equation*}
 These correspond to the entries for  $({}^{T}B^a)$,  $({}^{P}B^a)$, and  $({}^{TP}B^a)$ in the final row of the table.\footnote{Here we have worked with what might be called the ``invariant angular velocity" of $\xi^a$.  If we worked, instead, with the angular velocity of $\xi^a$ as determined relative to the frame $\eta^a$, the formal parallel with $B^a$ would be even closer. The later ``relativized angular velocity" is defined by: 
 \[
\hat{\omega}^a  =   \frac{1}{2} \, \epsilon^{acd} \, \nabla_c  \, \xi_d =  \frac{1}{2} \, \epsilon^{abcd} \, \eta_b \, \nabla_c  \, \xi_d. 
\]
Here the dependence on $\epsilon_{abc}$, characteristic of \emph{axial} vector fields,  is explicit. (To see where this expression comes from, notice that if $\hat{\xi}^a$ is  $h^a_{\ b} \, \xi^b$, the field that  results from projecting $\xi^a$ orthogonal  to $\eta^a$, then the curl of $\hat{\xi}^a$ (relative to $\eta^a$) is $\frac{1}{2} \, \epsilon^{acd} \, \nabla_c  \, \hat{\xi}_d$. But, since $h^a_{\ b}$ is constant, and $\epsilon^{acd} \, \eta_d = \epsilon^{abcd} \, \eta_b \,  \eta_d = \mathbf{0}$,   
\[
\frac{1}{2} \, \epsilon^{acd} \, \nabla_c  \, \hat{\xi}_d = \frac{1}{2} \, \epsilon^{acd} \, h_d^{\ n} \,  \nabla_c  \, \xi_n =
\frac{1}{2} \, \epsilon^{acd} \, (g_d^{\ n} - \eta_d \, \eta^n) \,  \nabla_c  \, \xi_n = \frac{1}{2} \, \epsilon^{acd} \, \nabla_c  \, \xi_d.
\] 
So  $\hat{\omega}^a$ is just the curl of $\hat{\xi}^a$.) $\hat{\omega}^a$ transforms exactly like $\omega^a$ under the operations  $\mathbf{T}$, $\mathbf{P}$, and $\mathbf{TP}$.}
 
Think about it this way. If we make a movie of a fluid whirling in a clockwise direction, and then play the movie backwards, we see the fluid whirling in a counterclockwise direction.  The angular velocity of the fluid is reversed.  On the other hand, if we play it backwards, project the image onto a mirror, and  then watch the reflected image, we see the fluid whirling in a clockwise direction again, as in the original. In this case, the angular velocity is not reversed.
 
 I can imagine David Albert objecting that the proposed analogy between a magnetic field and an angular velocity field  is misleading. Presumably, he would agree that the latter (understood as a component of a fluid's instantaneous dynamical condition) is properly reversed by the time reversal operation, just as is the linear velocity of a point particle. (Recall the  discussion in section 2.) But he would continue to insist that ``magnetic fields are not -- either logically or conceptually -- the \emph{rates of change} of anything", and so should not be reversed.
 
 Of course, I do not think the response is to the point. Magnetic fields may not be rates of change of anything in the appropriate sense, but they \emph{are} axial vector fields.  The claim here is that one does not \emph{have} a magnetic field $B^a$ until  a spatial orientation is specified (in addition to a temporal orientation). And, similarly, one does not \emph{have} a time reversed field $({}^TB^a)$ until a spatial orientation is specified (in addition to the reversed temporal orientation). At issue is whether the second spatial orientation is taken to be the same as the first or not i.e., whether we are considering the action of \textbf{T} on $B^a$ or, rather, \textbf{TP}. The first option leads to the standard account of how the time reversal operator acts on magnetic fields; the second leads to Albert's account.  One wants to say to him that the time reversal operation, properly understood, simply does not -- ``either logically or conceptually"-- include spatial parity reversal as a constituent sub-operation.
 \section{Conclusion}
I have tried to show that David Albert's claims concerning the (non) time reversal invariance of classical electromagnetic theory are based on a misleading way of thinking about the representation of electromagnetic  fields, and that some insight is achieved into issues of time reversal invariance when one thinks about them in terms of four-dimensional spacetime geometry. I have argued (section 6) that the inversion of magnetic fields under time reversal that so troubles Albert is benign and makes good geometric sense. I have also argued (section 4) that one can formulate and defend the claim that classical electromagnetic theory is time reversal invariant without making reference to magnetic fields at all.  

Defending orthodoxy is much less exciting work than embracing heresy. But it has its place on occasion. I  hope I have succeeded in restoring the faith of at least a few of my colleagues. 

\section*{Acknowledgements}
I wish to thank Frank Arntzenius, Jordi Cat, Erik Curiel, Dennis Dieks, Robert Geroch, Tim Maudlin, John Norton, Howard Stein, Jos Uffink, and an anonymous referee for helpful comments. (They certainly do not all agree with the claims I have made.)

\section*{References}
\begin{description}
\item []  Albert, D. Z. (2000).  \emph{Time and chance}.  Cambridge (MA), London:  Harvard University Press.
\vspace{-2.2em}
\item [] Arntzenius, F. (2000).  Are there really instantaneous velocities?  \emph{The Monist}, \emph{83}, 187-208.
\vspace{-.5em}
\item [] Arntzenius, F. (2003/4?). Time reversal operations and the direction of time. Forthcoming in \emph{Studies in the History and Philosophy of Modern Physics}.
\vspace{-.5em}
 \item []  Earman, J. (2002).  What time reversal invariance is and why it matters. \emph{International Studies in the Philosophy of Science}, \emph{16}, 245-264.
 \vspace{-.5em}
\item []  Floyd, S. (2003).  On the possibility of instantaneous velocities.  unpublished. 
\vspace{-.5em}
 \item [] Smith, S. (2003).  Are instantaneous velocities real and really instantaneous?:  an argument for the affirmative. 
\emph{Studies in the History and Philosophy of Modern Physics}, \emph{34}, 261-280. 
 \vspace{-.5em}
 \item []  Wald, R.  M. (1984). \emph{General relativity}. Chicago:  University of Chicago Press.
\end{description} 

\end{document}